\documentclass[12pt]{article}
\usepackage{amsthm}

\theoremstyle{definition}
\newtheorem{definition}{Definition}

\title{A Data Mining Framework for Optimal Product Selection in Retail Supermarket Data: The Generalized PROFSET Model}

\author{Tom Brijs\footnote{Tom Brijs is a research fellow of the Fund for Scientific Research Flanders.} 
\quad Bart Goethals \\
Gilbert Swinnen \quad Koen Vanhoof \quad Geert Wets \\
University of Limburg
}

\date{}

\begin{document}
\maketitle

\begin{abstract}
In recent years, data mining researchers have developed efficient
association rule algorithms for retail market basket analysis.
Still, retailers often complain about how to adopt association
rules to optimize concrete retail marketing-mix decisions. It is
in this context that, in a previous paper, the authors have
introduced a product selection model called
PROFSET.\footnote{PROFSET stands for PROFitability per SET because
the optimization model is based on the calculation of the
profitability per frequent set in order to determine the
cross-selling potential between products.} This model selects the
most interesting products from a product assortment based on
their cross-selling potential given some retailer defined
constraints.  However this model suffered from an important
deficiency: it could not deal effectively with supermarket data,
and no provisions were taken to include retail category
management principles.  Therefore, in this paper, the authors
present an important generalization of the existing model in
order to make it suitable for supermarket data as well, and to
enable retailers to add category restrictions to the model.
Experiments on real world data obtained from a Belgian
supermarket chain produce very promising results and demonstrate
the effectiveness of the generalized PROFSET model.
\end{abstract}

\section{Introduction}

Since almost all mid to large size retailers today possess
electronic sales transaction systems, retailers realize that
competitive advantage will no longer be achieved by the mere use
of these systems for purposes of inventory management or
facilitating customer check-out. In contrast, competitive
advantage will be gained by those retailers who are able to
extract the knowledge hidden in the data, generated by those
systems, and use it to optimize their marketing decision making.
In this context, knowledge about how customers are using the
retail store is of critical importance and distinctive
competencies will be built by those retailers who best succeed in
extracting actionable knowledge from these data.  Association rule
mining \cite{ais} can help retailers to efficiently extract this
knowledge from large retail databases.  We assume some
familiarity with the basic notions of association rule mining.

In recent years, a lot of effort in the area of retail market
basket analysis has been invested in the development of
techniques to increase the interestingness of association rules.
Currently, in essence three different research tracks to study
the interestingness of association rules can be distinguished.

First, a number of objective measures of interestingness have
been developed in order to filter out non-interesting association
rules based on a number of statistical properties of the rules,
such as support and confidence \cite{ais}, interest
\cite{correlation}, intensity of implication \cite{implic},
J-measure \cite{nar}, and correlation \cite{prune}. Other
measures are based on the syntactical properties of the rules
\cite{p_analysis}, or they are used to discover the
least-redundant set of rules \cite{redundancy}. Second, it was
recognized that domain knowledge may also play an important role
in determining the interestingness of association rules.
Therefore, a number of subjective measures of interestingness
have been put forward, such as unexpectedness
\cite{unexpectedness}, actionability \cite{actipat} and rule
templates \cite{interest}. Finally, the most recent stream of
research advocates the evaluation of the interestingness of
associations in the light of the micro-economic framework of the
retailer \cite{papadimitriou}. More specifically, a pattern in
the data is considered interesting only to the extent in which it
can be used in the decision-making process of the enterprise to
increase its utility.

It is in this latter stream of research that the authors have
previously developed a model for product selection called PROFSET
\cite{profset}, that takes into account both quantitative and
qualitative elements of retail domain knowledge in order to
determine the set of products that yields maximum cross-selling
profits. The key idea of the model is that products should not be
selected based on their individual profitability, but rather on
the \emph{total} profitability that they generate, including
profits from cross-selling. However, in its previous form, one
major drawback of the model was its inability to deal with
supermarket data (i.e., large baskets). To overcome this
limitation, in this paper we will propose an important
generalization of the existing PROFSET model that will
effectively deal with large baskets.  Furthermore, we generalize
the model to include category management principles specified by
the retailer in order to make the output of the model even more
realistic.

The remainder of the paper is organized as follows.  In
Section~\ref{overview} we will focus on the limitations of the
previous PROFSET model for product selection.  In
Section~\ref{general}, we will introduce the generalized PROFSET
model.  Section~\ref{impl} will be devoted to the empirical
implementation of the model and its results on real-world
supermarket data.  Finally, Section~\ref{concl} will be reserved
for conclusions and further research.

\section{The PROFSET Model} \label{overview}

The key idea of the PROFSET model is that when evaluating the
business value of a product, one should not only look at the
individual profits generated by that product (the na\"{i}ve
approach), but one must also take into account the profits due to
cross-selling effects with other products in the assortment.
Therefore, to evaluate product profitability, it is essential to
look at frequent sets rather than at individual product items
since the former represent frequently co-occuring product
combinations in the market baskets of the customer. As was also
stressed by Cabena et al.\ \cite{cabena}, one disadvantage of
associations discovery is that there is no provision for taking
into account the business value of an association.  The PROFSET
model was a first attempt to solve this problem. Indeed, in terms
of the associations discovered, the sale of an expensive bottle
of wine with oysters accounts for as much as the sale of a carton
of milk with cereal.  This example illustrates that, when
evaluating the interestingness of associations, the
micro-economic framework of the retailer should be incorporated.
PROFSET was developed to maximize cross-selling opportunities by
evaluating the profit margin generated per frequent set of
products, rather than per product.  In the next Section we will
discuss the limitations of the previous PROFSET model.  More
details can be found elsewhere \cite{profset}.

\subsection{Limitations} \label{limit}

The previous PROFSET model was specifically developed for market
basket data from automated convenience stores.  Data sets of this
origin are characterized by small market baskets (size 2 or 3)
because customers typically do not purchase many items during a
single shopping visit. Therefore, the profit margin generated per
frequent purchase combination $(X)$ could accurately be
approximated by adding the profit margins of the market baskets
$(T_j)$ containing the same set of items, i.e.\, $X =T_j$.
However, for supermarket data, the existing formulation of the
PROFSET model poses significant problems since the size of market
baskets typically exceeds the size of frequent itemsets. Indeed,
in supermarket data, frequent itemsets mostly do not contain more
than 7 different products, whereas the size of the average market
basket is typically 10 to 15.  As a result, the existing profit
allocation heuristic cannot be used anymore since it would cause
the model to heavily underestimate the profit potential from
cross-selling effects between products.  However, getting rid of
this heuristic is not trivial and it will be discussed in detail
in Section~\ref{profalloc}.

A second limitation of the existing PROFSET model relates to
principles of category management. Indeed, there is an increasing
trend in retailing to manage product categories as separate
strategic business units \cite{sagit}.  In other words, because
of the trend to offer more products, retailers can no longer
evaluate and manage each product individually.  Instead, they
define product categories and define marketing actions (such as
promotions or store layout) on the level of these categories. The
generalized PROFSET model takes this domain knowledge into
account and therefore offers the retailer the ability to specify
product categories and place restrictions on them.

\section{The Generalized PROFSET Model} \label{general}

In this section, we will highlight the improvements being made to
the previous PROFSET model \cite{profset}.

\subsection{Profit Allocation} \label{profalloc}

Avoiding the equality constraint $X = T_j$ results in different
possible profit allocation systems.  Indeed, it is important to
recognize that the margin of transaction $T_j$ can potentially be
allocated to different frequent subsets of that transaction.  In
other words, how should the margin $m(T_j)$ be allocated to one
or more different frequent subsets of $T_j$?

The idea here is that we would like to know the purchase
intentions of the customer who bought $T_j$. Unfortunately, since
the customer has already left the store, we do not possess this
information.  However, if we can assume that some items occur
more frequently together than others because they are considered
complementary by customers, then frequent itemsets may be
interpreted as purchase intentions of customers. Consequently,
there is the additional problem of finding out which and how many
purchase intentions are represented in a particular transaction
$T_j$.  Indeed, a transaction may contain several frequent subsets
of different sizes, so it is not straightforward to determine
which frequent sets represent the underlying purchase intentions
of the customer at the time of shopping.  Before proposing a
solution to this problem, we will first define the concept of a
maximal frequent subset of a transaction.

\begin{definition}
Let $F$ be the collection of all frequent subsets of a sales
transaction $T_j$. Then $X \in F$ is called \emph{maximal},
denoted as $X_{\it max}$, if and only if $\forall Y \in F : |Y|
\leq |X|$.
\end{definition}

Using this definition, we will adopt the following rationale to
allocate the margin $m(T_j)$ of a sales transaction $T_j$.

If there exists a frequent set $X = T_j$, then we allocate
$m(T_j)$ to $M(X)$, just as in the previous PROFSET model.
However, if there is no such frequent set, then one maximal
frequent subset $X$ will be drawn from all maximal frequent
subsets according to the probability distribution $\Theta_{T_j}$,
with
$$\Theta_{T_j}(X_{\it max}) = \frac{\mbox{support}(X_{max})}{\sum_{Y_{\it max} \in T_j} \mbox{support}(Y_{\it max})}$$
After this, the margin $m(X)$ is assigned to $M(X)$ and the
process is repeated for $T_j \setminus X$.  In summary:
\begin{center}
\parbox{\columnwidth}{
\begin{tabbing}
\quad \= \quad \= \kill
\textbf{for} every transaction $T_j$  \textbf{do} \{ \+ \\
 \textbf{while} ($T_j$ contains frequent sets) \textbf{do} \{ \+ \\
 Draw X from all maximal frequent subsets \\
 using probability distribution $\Theta_{T_j}$; \\ [\medskipamount]
 $M(X) := M(X) + m(X)$ \\
 with $m(X)$ the profit margin of $X$ in $T_j$; \\  [\medskipamount]
 $T_j := T_j \setminus X$; \- \\
\} \- \\
\} \\
\textbf{return} all $M(X)$;
\end{tabbing}
}
\end{center}
Say, during profit allocation, we are given a transaction 
$$T = \{\mbox{cola}, \mbox{peanuts}, \mbox{cheese}\}.$$
Table~\ref{subsets} contains all frequent subsets of $T$ for a
particular transaction da\-ta\-base.
\begin{table}
\centering \caption{Frequent Subsets of $T_{100}$} \label{subsets}
\begin{tabular}{|lccc|}
  \hline
  \textbf{Frequent Sets} & \textbf{Support} & \textbf{Maximal} & \textbf{Unique} \\
  \hline
  \{cola\} & $10\%$ & No & No \\
  \{peanuts\} & $5\%$ & No & No \\
  \{cheese\} & $8\%$ & No & No \\
  \{cola, peanuts\} & $2\%$ & Yes & No \\
  \{peanuts, cheese\} & $1\%$ & Yes & No \\ \hline
\end{tabular}
\end{table}
In this example, there is no \emph{unique} maximal frequent subset
of $T$.  Indeed, there are two maximal frequent subsets of $T$,
namely \{cola, peanuts\} and \{peanuts, cheese\}. Consequently,
it is not obvious to which maximal frequent subset the profit
margin $m(T)$ should be allocated.  Moreover, we would not
allocate the entire profit margin $m(T)$ to the selected itemset,
but rather the proportion $m(X)$ that corresponds to the items
contained in the selected maximal subset.

Now how can one determine to which of both frequent subsets of
$T$ this margin should be allocated?  As we have already
discussed, the crucial idea here is that it really depends on
what has been the purchase intentions of the customer who
purchased $T$. Unfortunately, one can never know exactly since we
haven't asked the customer at the time of purchase. However, the
support of the frequent subsets of $T$ may provide some
probabilistic estimation.  Indeed, if the support of a frequent
subset is an indicator for the probability of occurrence of this
purchase combination, then according to the data, customers buy
the maximal subset \{cola, peanuts\} two times more frequently
than the maximal subset \{peanuts, cheese\}. Consequently, we can
say that it is more likely that the customer's purchase intention
has been \{cola, peanuts\} instead of \{peanuts, cheese\}. This
information is used to construct the probability distribution
$\Theta_{T_j}$, reflecting the relative frequencies of the
frequent subsets of $T$.  Now, each time a sales transaction
\{cola, peanuts, cheese\} is encountered in the data, a random
draw from the probability distribution $\Theta_{T_j}$ will
provide the \emph{most probable} purchase intention (i.e.
frequent subset) for that transaction. Consequently, on average
in two of the three times this transaction is encountered,
maximal subset \{cola, peanuts\} will be selected and
$m(\{\mbox{cola}, \mbox{peanuts}\})$ will be allocated to
$M(\{\mbox{cola}, \mbox{peanuts}\})$. After this, $T$ is split up
as follows: $T := T \setminus \{\mbox{cola}, \mbox{peanuts}\}$
and the process of assigning the remaining margin is repeated as
if the new $T$ were a separate transaction, until $T$ does not
contain a frequent set anymore.

\subsection{Category Management Restrictions} \label{category}

As pointed out in Section~\ref{limit}, a second limitation of the
previous PROFSET model is its inability to include category
management restrictions.  This sometimes causes the model to
exclude even all products from one or more categories because
they do not contribute enough to the overall profitability of the
optimal set.  This often contradicts with the mission of
retailers to offer customers a wide range of products, even if
some of those categories or products are not profitable enough.
Indeed, customers expect supermarkets to carry a wide variety of
products and cutting away categories/departments would be against
the customers' expectations about the supermarket and would harm
the store's image.  Therefore, we want to offer the retailer the
ability to include category restrictions into the generalized
PROFSET model.

This can be accomplished by adding an additional index $k$ to the
product variable $Q_i$ to account for category membership, and by
adding constraints on the category level. Several kinds of
category restrictions can be introduced: which and how many
categories should be included in the optimal set, or how many
products from each category should be included.  The relevance of
these restrictions can be illustrated by the following common
practices in retailing.  First, when composing a promotion
leaflet, there is only limited space to display products and
therefore it is important to optimize the product composition in
order to maximize cross-selling effects between products and
avoid product cannibalization.  Moreover, according to the
particular retail environment, the retailer will include or
exclude specific products or product categories in the leaflet.
For example, the supermarket in this study attempts to
differentiate from the competition by the following image
components: \emph{fresh}, \emph{profitable} and \emph{friendly}.
Therefore, the promotion leaflet of the retailer emphasizes
product categories that support this image, such as fresh
vegetables and meat, freshly-baked bread, ready-made meals, and
others. Second, product category constraints may reflect shelf
space allocations to products.  For instance, large categories
have more product facings than smaller categories.  These kind of
constraints can easily be included in the generalized PROFSET
model as will be discussed hereafter.

\subsection{The Generalized PROFSET Model}

Bundling the improvements suggested in Sections~\ref{profalloc}
and~\ref{category} results in the generalized PROFSET model as
presented below.

Let categories $C_1, \ldots, C_n$ be sets of items, $L$ the set
of frequent itemsets, and let $P_X$, $Q_i \in \{0,1\}$ be the
decision variables for which the optimization routine must find
the optimal values. $P_X$ specifies whether an itemset $X$ will
positively contribute to the value of the objective function, and
$Q_i$ equals 1 as soon as any itemset $X$ in which it is included
is set to 1 ($P_X = 1$) by the optimization routine. Let ${\rm
Cost}_i$ be the inventory and handling cost of item $i$. The
objective of the following formula is to maximize all profits
from cross-selling effects between products:

$$\mbox{max}\left( \sum_{X \in L} M(X) P_X - \sum_{c=1}^{n}\sum_{i \in C_c} {\rm Cost}_i Q_i\right)$$

which is subject to the following constraints
\begin{eqnarray}
\label{c1}
\sum_{c=1}^{n} \sum_{i \in C_c} Q_i = \mbox{ItemMax} \\
\label{c2}
\forall X \in L,\:\forall i \in X:Q_i \geq P_X \\
\label{c3} \forall C_c: \sum_{i \in C_c} Q_i \geq
\mbox{ItemMin}_{C_c}
\end{eqnarray}

Constraint~\ref{c1} determines how many items are allowed to be
included in the optimal set. The $\mbox{ItemMax}$ parameter,
specified by the retailer, will depend on the retail environment
in which the model is being used.  For instance, it may be the
number of eye-catchers (products obtaining special display space)
in the supermarket or the number of facings in a promotion
leaflet. Constraint~\ref{c2} is analogous to the one in the
previous PROFSET model and specifies the relationship between the
frequent sets and the products contained in them. Finally,
constraint~\ref{c3} specifies the number of categories and the
number of products that are allowed, within each category, to
enter the optimal set.

\section{Empirical Study} \label{impl}

The empirical study is based on a data set of $18\,182$ market
baskets obtained from a sales outlet of a Belgian supermarket
chain over a period of 1 month.  The store carries $9\,965$
different products grouped in 281 product categories.  The
average market basket contains $10.6$ different product items.  In
total, $3\,381$ customers own a loyalty card of the supermarket under
study.

First, frequent sets and association rules were discovered from
the market baskets with a minimum absolute support threshold of
30 transactions. The motivation behind this is that a product or
set of products should have been sold at least, approximately,
once a day to be called frequent. Slightly more than $87\%$ of
the products are sold less than once a day.

The retailer in question is interested in finding the optimal set
of eye-catchers such that the profit from cross-selling these
eye-catchers is maximized. Hence, this should be represented by
the objective function as described in the previous section.
However, because of limited shelf-space for each product
category, the retailer specified that each product category can
only delegate one product to the optimal set, represented by the
category constraint (i.e. constraint~\ref{c3}). Subsequently, it
is the goal of the generalized PROFSET model to select the most
profitable set of products in terms of cross-selling
opportunities between the delegates of each category.

For 54 $(24,7\%)$ of the 218 product categories, the generalized
PROFSET model selects a different product than the one with the
highest individual profit ranking within each category.  This
suggests that for these products, there must be some
cross-selling opportunity with eye-catchers from other categories
which cause these products to get \emph{promoted} in the
profitability ranking.

Due to space limitations Table~\ref{profit} shows the relative
improvements in cross-selling profit for only some categories,
expressed as the percentage of improvement in cross-selling
profits by choosing the optimal products from the generalized
PROFSET model instead of selecting the product with the highest
individual profitability within each category.

\begin{table}
\caption{Cross-selling profit improvements} \label{profit}
\centering
\begin{tabular}{|lc|}
  \hline
  \textbf{Category} & \textbf{Improvement} \\
  \hline
Washing-up liquid & 21\% \\
Baby food & 49\% \\
Margarine 1 & 189\% \\
Coffee biscuits & 14\% \\
Sandwich filling & 43\% \\
Candy bars & 588\% \\
Canned fish & N/A \\
Canned fruit & 3\% \\
Packed-up bread & 8\% \\
Newspapers and magazines & 55\% \\
$\ldots$ & $\ldots$ \\
\hline
\end{tabular}
\end{table}
\begin{table*}[t]
\centering \caption{Own and cross-selling profit figures (in BEF)
per product} \label{cross}
\begin{tabular}{|lccc|}
  \hline
  & \textbf{Own} & \textbf{Cross-selling} & \textbf{Total} \\
  \textbf{Product} & \textbf{profit} & \textbf{profit} & \textbf{profit} \\
  \hline
  1. {\sc milky way mini} & $37\,808$ & $2\,350$ & $40\,158$ \\
  2. {\sc melo cakes} & $34\,333$ & 0 & $34\,333$ \\
  3. {\sc Leo 3-pack} & $28\,728$ & 0 & $28\,728$ \\
  4. {\sc Leo 10-pack} 10+2 & $12\,028$ & $264\,228$ & $276\,256$ \\
\hline
\end{tabular}
\end{table*}

It would lead us too far to discuss the profit improvements in
detail for all categories.  Therefore, we will highlight one of
the most striking results to illustrate the power of the model.
Analogous conclusions can be obtained for other categories.  Note
that N/A means that there is no alternative product available in
that category that has enough support to be frequent, such that
comparison with the product, selected by the generalized PROFSET
model, is not applicable. For instance, for the category candy
bars, the profit from cross-selling the selected eye-catcher of
this category with eye-catchers of other categories would increase
cross-selling profits by $588\%$. This can be observed in
Table~\ref{cross} (only relevant products are included).

Table~\ref{cross} illustrates that product 4 in the candy bars
category is ranked last when looking at its own profit. However,
due to large cross-selling effects with eye-catchers of other
product categories, this product becomes much more important when
looking at the total profit. This illustrates that for the
eye-catchers application, it is better to display product `Leo
10-pack 10+2' than to display one of its competing products in
the same category. In contrast, if the objective would be the
selling volume of the individual product, then it would be better
to select product 1 as eye-catcher, but since the retailer wants
the customer to buy other products with it, product 4 will
definitely be the best choice.  The association rules discovered
during the mining phase validate these conclusions. \\
[\medskipamount]
{\sc milky way}$\Rightarrow${\sc vegetable/fruit} \\
(sup=$0.17\%$, conf=$50.82\%$) \\
{\sc meat product and Leo 10-pack} $\Rightarrow${\sc cheese
product} \\
(sup=$0.396\%$, conf=$55\%$)
\\ [\medskipamount] Note that the products included in the rules
are all eye-catchers such as determined by the generalized
PROFSET model.  The reason that the other items contained in the
association rules carry a rather abstract name, such as ``cheese
product'', is because this is a collective noun for products that
do not have an own barcode, like for instance different cheese
products that are weighed at the check-out after which they are
grouped into an abstract product name such as ``cheese product''.

Finally, for those product categories that do not contain
frequent products, the generalized PROFSET model will choose the
product with the highest individual profit in order to maximize
the overall profitability of the eye-catcher set.

\section{Further Research} \label{concl}

The authors plan to test the proposed model in practice and
externally validate its performance based on a real world
experiment in cooperation with the Belgian supermarket chain.
Furthermore, additional improvements to the model will be
considered.  More specifically, it will be studied how promotion
coupons affect the composition of the optimal set of products and
whether it is possible to measure the effect of the value price
reduction on the cross-selling profitability of products.

\bibliographystyle{plain}

\end{document}